\documentclass[12pt,preprint]{emulateapj}
\usepackage[utf8]{inputenc}
\usepackage[normalem]{ulem}
\usepackage{color}

\begin{document}
\slugcomment{{\sc Accepted to AJ:} March 2, 2017}
\title{Transiting Planets with LSST III:  Detection Rate per Year of Operation}
\author{Savannah R. Jacklin\altaffilmark{1,2}, 
Michael B.\ Lund\altaffilmark{2}, 
Joshua Pepper\altaffilmark{3}, 
Keivan G.\ Stassun\altaffilmark{2,1}}
\altaffiltext{1}{Department of Physics, Fisk University, Nashville, TN 37208, USA}
\altaffiltext{2}{Department of Physics and Astronomy, Vanderbilt University, Nashville, TN 37235, USA}
\altaffiltext{3}{Department of Physics, Lehigh University, Bethlehem, PA 18015, USA}

\begin{abstract}
The Large Synoptic Survey Telescope (LSST) will generate light curves for approximately 1~billion stars.  Our previous work has demonstrated that, by the end of the LSST 10-yr mission, large numbers of transiting exoplanetary systems could be recovered using the LSST ``deep drilling" cadence.  Here we extend our previous work to examine how the recoverability of transiting planets over a range of orbital periods and radii evolves per year of LSST operation.  As specific example systems we consider hot Jupiters orbiting solar-type stars and hot Neptunes orbiting K-Dwarfs at distances from Earth of several kpc, as well as super-Earths orbiting nearby low-mass M-dwarfs.  The detection of transiting planets increases steadily with the accumulation of data over time, generally becoming large ($\gtrsim$10\%) after 4--6 yr of operation. However, we also find that short-period ($\lesssim$2~d) hot Jupiters orbiting G-dwarfs and hot Neptunes orbiting K-dwarfs can already be discovered within the first 1--2 yr of LSST operation.
\end{abstract}

\section{Introduction\label{sec:intro}}
In the era of large all sky survey telescopes such as LSST, it has become increasingly important for astronomers to develop the appropriate tools to analyze the veritable flood of data that will soon be available. 
LSST is currently under construction on Cerro Pachón in Chile, and after a scheduled first light in $\sim2020$, LSST will take approximately 30 terabytes of data per night over nearly the entire southern sky \citep{Ivezic:2008}.  As a wide-fast-deep survey, LSST is built to satisfy a wide variety of science goals over an unprecedented volume of space.  Per \citet{Ivezic:2008} the mission has four primary science goals: understanding dark matter and dark energy, cataloging the solar system, Milky Way structure and formation, and the study of transient and variable objects (i.e. the changing sky).  This paper examines the potential use of LSST for detection of transiting exoplanets and falls into the lattermost category.

Unlike missions such as {\it Kepler} \citep{Borucki:2010} and TESS \citep{Ricker:2015}, or small, ground-based surveys such as KELT \citep{Pepper:2007,Pepper:2012}, HATNet \citep{Bakos:2004}, or SuperWASP \citep{Pollacco:2006}, LSST is not optimized for exoplanet detection.  LSST will operate at two primary cadences, hereafter referred to as ``regular" and ``deep-drilling".  Regular cadence observations will constitute approximately 90\% of the telescope's observation time, with target objects receiving approximately 1000 observations over the course of LSST's ten year mission.  Deep drilling cadence observations will take up the remaining 10\% of LSST's observation time, with target objects receiving $\sim10,000$ observations after ten years \citep{Ivezic:2008}.  Although these cadences are not ideal for exoplanet detection, we showed in \citet{Lund:2015} and \citet{Jacklin:2015} that the sheer number of LSST targets ($\sim10^{9}$) makes a large number of hot Jupiter and other short-period exoplanet detections likely, particularly in the deep-drilling mode.

Indeed, our previous work has demonstrated that LSST will have the ability to detect exoplanets in many interesting regions of parameter space, and because of its magnitude depth and sky coverage will include types of systems that are at present poorly explored. This includes planets orbiting nearby late-type stars and larger planets orbiting Sun-like stars at very large distances.  While \citet{Lund:2015} and \citet{Jacklin:2015} considered the yield of transiting exoplanets for the full 10-yr LSST dataset, it is worth investigating
how quickly LSST is likely to achieve various types of exoplanet discoveries over the course of its lifetime.  Using the LSST Operation Simulation (OpSim) of deep-drilling fields, in this paper we predict the rate of transiting exoplanet detection by LSST as a function of year of observation.  

The paper proceeds as follows:  In Section~\ref{sec:methods} we discuss our strategy for generating simulated light curves, the fiducial star-planet systems simulated, the method of recovering the period of transits, and checking the recovery results for false positives in order to determine detectability.  Section~\ref{sec:results} presents the results for the detectability of various types of transiting planets, as a function of time throughout the LSST 10-year mission, for each of the fiducial stellar systems.  Our results show that LSST will be able to detect a wide variety of exoplanets orbiting several different types of stars at periods ranging from 0.5--20 days, and that it can do so within the first few years of data collection in the deep-drilling fields for very short-period planets.  We conclude with a brief discussion, including planned future work, and a summary in Section~\ref{sec:disc}.

\section{Methods}\label{sec:methods}

Here we describe the methods by which we generate simulated LSST light curves, list the fiducial transiting exoplanets that we will analyze throughout this work, describe how we inject simulated transits of these exoplanets into the simulated light curves, and describe the period-search methods we use to probe the detection of the transits in the simulated light curves.

\subsection{Simulated Light Curve Generation}\label{sec:lightcurves}

We create simulated light curves using the method described in \citet{Lund:2015,Jacklin:2015} which we summarize here.  Each light curve depends on five parameters: host star mass, radius of transiting planet, distance of the system from Earth, period of transiting planet, and LSST cadence (i.e. regular or deep-drilling).  Each of the systems simulated assumes circular orbits with equatorial transits (i.e., we do not consider the effects of orbital eccentricity or grazing transits).

All host stars in our simulations (see Section~\ref{sec:systems}) are dwarfs with spectral types determined via mass interpolation from Table 15.8 in \citet{Cox:2000}, absolute magnitudes per LSST filter from \citet{Covey:2007}, and radii derived using the mass-radius relationship from \citet{Beatty:2008}.  
With this information, we create continuous and noiseless light curves.  We then inject a simple boxcar transit into each of the light curves with a duration based on the planetary orbital period and stellar radius, and a depth dependent on the ratio of the stellar to planetary radii.  In this analysis we assume circular orbits with equatorial transits, as well as noiseless stellar hosts.

With continuous light curves for a range of stellar radii, spectral types, planet radii, and planet periods (see Section~\ref{sec:systems}), we time-sample them using the LSST operation simulation (OpSim) v2.3.2 run 3.61.  The OpSim simulates ten full years of LSST observations considering factors such as weather variations and downtime for telescope maintenance.  Out of the many parameters available from OpSim, we utilize time of observation, filter, and limiting magnitude.  Based on our previous findings that the vast majority of transiting exoplanet detections will occur in the deep-drilling fields \citep{Lund:2015,Jacklin:2015}, we therefore consider only the deep-drilling cadence from the OpSim. 

Once a light curve has been simulated we sample it and place our planet-star systems at a chosen distance and calculate the star's apparent magnitude in LSST's $g, r,$ and $i$ bands, which exhibit the least random photometric noise as shown in \citet{Lund:2015}.  We then calculate the expected total per-visit photometric precision per LSST band using the stellar apparent magnitude with:
\begin{equation}
\sigma_{tot}^{2} = \sigma_{sys}^{2} + \sigma_{rand}^{2} \,\,,
\end{equation}
as described in Ivezic et al. (2008).  We take $\sigma_{sys}$ as the system designed noise floor of 0.005 for each band.  The random photometric noise $\sigma_{rand}$ varies with respect to a band-specific parameter ($\gamma$) and the apparent magnitude, and is calculated by:
\begin{equation}
\sigma_{rand}^{2} = (0.04 - \gamma)x + \gamma x^{2}\,\,,
\end{equation}
in which $x = 10^{(m - m_{5})}$ where $m_{5}$ is the $5\sigma$ limiting magnitude per LSST filter.  The value of $m_{5}$ is generated by OPSIM and changes each visit based on sky brightness, seeing, exposure time, airmass, atmospheric extinction, and instrument throughput \citep{Ivezic:2008}.  We add noise to the light curves on a per-band basis in order to simulate observed S/N.  After noise is added, the $g$, $r$, and $i$ band light curves are median-subtracted and combined in order to form one master light curve.  Table \ref{tab:table1} lists the absolute magnitude, total magnitude, and total noise as a function of LSST filter and band for each of our simulated systems described in Section \ref{sec:systems}.

We note that this treatment of noise in the light curves does not at present include the possible effects of contamination by neighboring stars, which could become important for crowded regions such as the Galactic plane, bulge, and possibly also the Magellanic Clouds. We defer a treatment of this additional complication to a followup paper (Lund et al., in preparation), and for our current purposes emphasize that the approach laid out here applies, strictly speaking, to non-crowded fields.

We also do not attempt to simulate the possible effects of stellar noise (e.g., activity). As described in the following section, we have selected transiting planets to simulate that would result in relatively large transit signals of $\gtrsim$1\%, which should not be significantly affected by typical activity levels on most solar-type stars \citep[see, e.g.,][]{Basri:2011}. Typical activity levels on M-dwarfs could be more important. For the purposes of this work we simply caution that our simulations of transiting planets for M-dwarfs apply in the case of relatively inactive stellar hosts.

\begin{table}[]
    \centering 
    \caption{Stellar Noise}
    \begin{tabular}{ccccc}
    \hline \hline
        Mass ($M_{\odot}$) & Filter & Abs Mag & App Mag & Total Noise (mag)\\ \hline
        1.0 & g  & 5.84 & 20.07 & 0.01\\
        1.0 & r &  4.47 & 18.70 & 0.01\\
        1.0 & i & 4.31 & 18.54 & 0.01\\
        \hline
        0.6 & g & 10.21 & 24.44 & 0.02\\
        0.6 & r & 7.65 & 21.87 & 0.02\\
        0.6 & i & 7.10 & 21.33 & 0.12\\
        \hline
        0.25 & g & 14.41 & 28.63 & 5.54 \\
        0.25 & r & 11.31 & 25.53 & 0.43\\
        0.25 & i & 9.72 & 23.94 & 0.19\\
        \hline
        \vspace{0.5mm}
    \end{tabular}
    {Absolute magnitude, total magnitude, and total noise as a function of LSST filter and band for each of simulated system.}
    \vspace{2mm}
    \label{tab:table1}
\end{table}

\subsection{Example Exoplanet Systems Simulated}\label{sec:systems}
\begin{figure}[!htb]
    \centering
    \includegraphics[width=\linewidth]{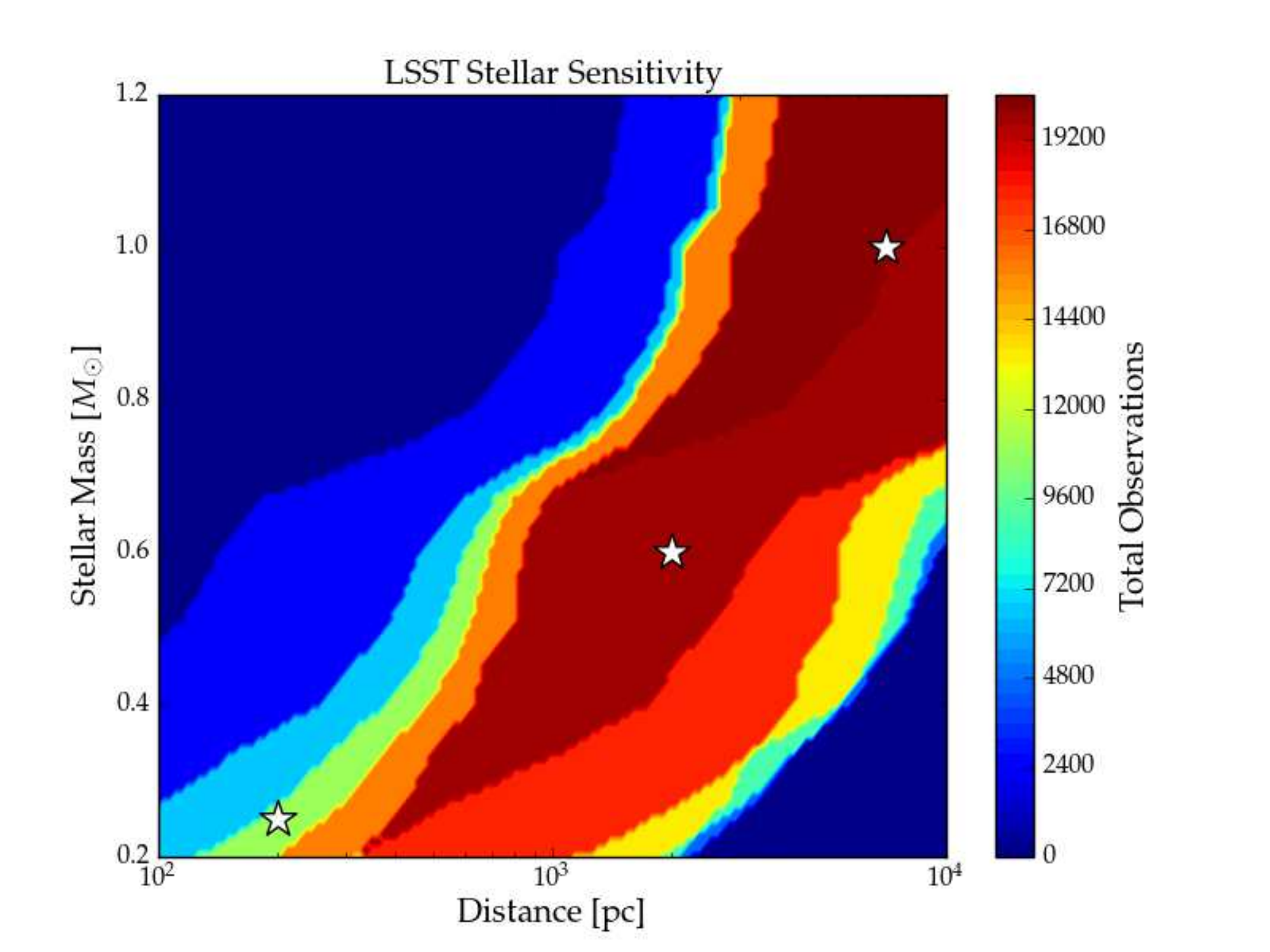}
    \caption{This figure displays, for a given stellar mass and distance, how many observations will be of sufficient precision ($<$ 30 mmag) in a DD field to permit exoplanet detection after 10 years of operation.  Warm colors indicate a large number of observations over several LSST bands, and cooler colors indicate fewer observations with sufficient precision. White stars indicate the stellar systems we explore in this paper.  There are $\sim$20,000 total deep-drilling observations in OpSim v2.3.2 run 3.61, although the typical numbers discussed in the past for the DD fields have been roughly 10,000 observations \citep{Ivezic:2008} due to the ongoing development of the instrument and simulation techniques.}
    \label{fig:systems}
\end{figure}

In order to explore a representative range of results, we simulate three fiducial exoplanet systems.  We begin with the same fiducial system as in our previous works \citep{Lund:2015,Jacklin:2015}, namely, a hot Jupiter with a radius of $10 R_{\oplus}$ orbiting a $1 M_{\odot}$ G-dwarf host star at 7~kpc.  Second, we simulate a hot Neptune with a radius of $6 R_{\oplus}$ orbiting a $0.6 M_{\odot}$ K-dwarf at 2~kpc. Third, we analyze a hot Super-Earth with a radius of $2 R_{\oplus}$ orbiting a $0.25 M_{\odot}$ M-dwarf at 200~pc.  Each of these cases was chosen specifically to fit within LSST's expected parameter space (i.e., within the detection and saturation limits of all filters; see Fig.~\ref{fig:systems}) \citep{Lund:2016}.  The planet radii were selected to create an approximate 1\% drop in total stellar flux as observed from Earth.

The quality of light curves generated for exoplanet detection primarily depends on the number of points included, which for LSST is $\sim$1000 points for objects in regular fields and $\sim$10000 for objects in deep drilling fields.  Depending on the brightness of the system's host star, some bands will not be useful for exoplanet detection (see \citet{Lund:2015, Jacklin:2015} for a full discussion on optimal bands), and therefore it is important to understand how many of the observations taken in a deep-drilling field will actually be useful.  We can parametrize detectability by photometric RMS precision, according to the absolute magnitude (determined by stellar mass) and the distance to the star, thus yielding the apparent magnitude.  This is then used to calculate expected RMS precision via the methods discussed in in ~\ref{sec:lightcurves}.  We then exclude all observations for a given star for which the precision is worse than 30 mmag, a fiducial precision that will be investigated in detail in later work.

We wish to examine what types of stars will be most suitable for transit searches in LSST deep drilling fields.  Figure~\ref{fig:systems} shows the number of observations that LSST will take, over 10 years, of a given object at deep drilling cadence (i.e., its sensitivity) where the noise is less than 30 mmag per observation.  The colored limits in Figure~\ref{fig:systems} are defined by the saturation and detection thresholds of LSST. 

%\newpage
\subsection{Recovery of Simulated Planets}

\subsubsection{Period Search with Box-fitting Least Squares}
To detect the planetary transits in the simulated light curves, we use the Box-fitting Least Squares (BLS) algorithm \citep{Kovacs:2002} as implemented through the VARTOOLS software package \citep{Hartman:2016}.  Specific details about our usage of BLS as a period finder are described in \citet{Lund:2015,Jacklin:2015}.  For this work, we perform the BLS period search in all cases over the period range $0.5 \le P \le 20$~d.  We implement BLS on each of our simulated light curves for each system architecture (see Section~\ref{sec:systems}).  To probe the likelihood of exoplanet detection with the accumulation of LSST data over time, we apply the BLS detection algorithm to each simulated light curve 10 times, once each for elapsed time $\Delta t = [1,10]$~yr from the start of LSST observations.

\subsubsection{Exoplanet Detection}

In this work, we deem an exoplanet ``recovered" if the top period returned by BLS is within 0.1\% of the input period.  An exoplanet is ``detected" if the period is recovered and the false positive probability is less than 0.1\%.  The false positive probability varies by the number of years of accumulated data.  

We calculate the false positive probability by creating an equivalent light curve with no injected transit for each cadence (i.e. regular or deep-drilling), stellar mass, distance, and year of observation.  The resultant light curves are then run through BLS, which calculates the highest power peak returned by a non-transiting system.  This process returns ten top values for the BLS powers, one for each year of observation, for a given stellar mass and distance combination.  These values of the BLS power are used as a comparison template for each transiting system sharing the same stellar mass and distance: if the period of a transiting planet is recovered, and the BLS power of that period is greater than the false alarm probability, the planet is considered detected.

\subsubsection{Mitigation of Diurnal Sampling}

\begin{figure}[!htb]
    \centering
    \includegraphics[width=\linewidth]{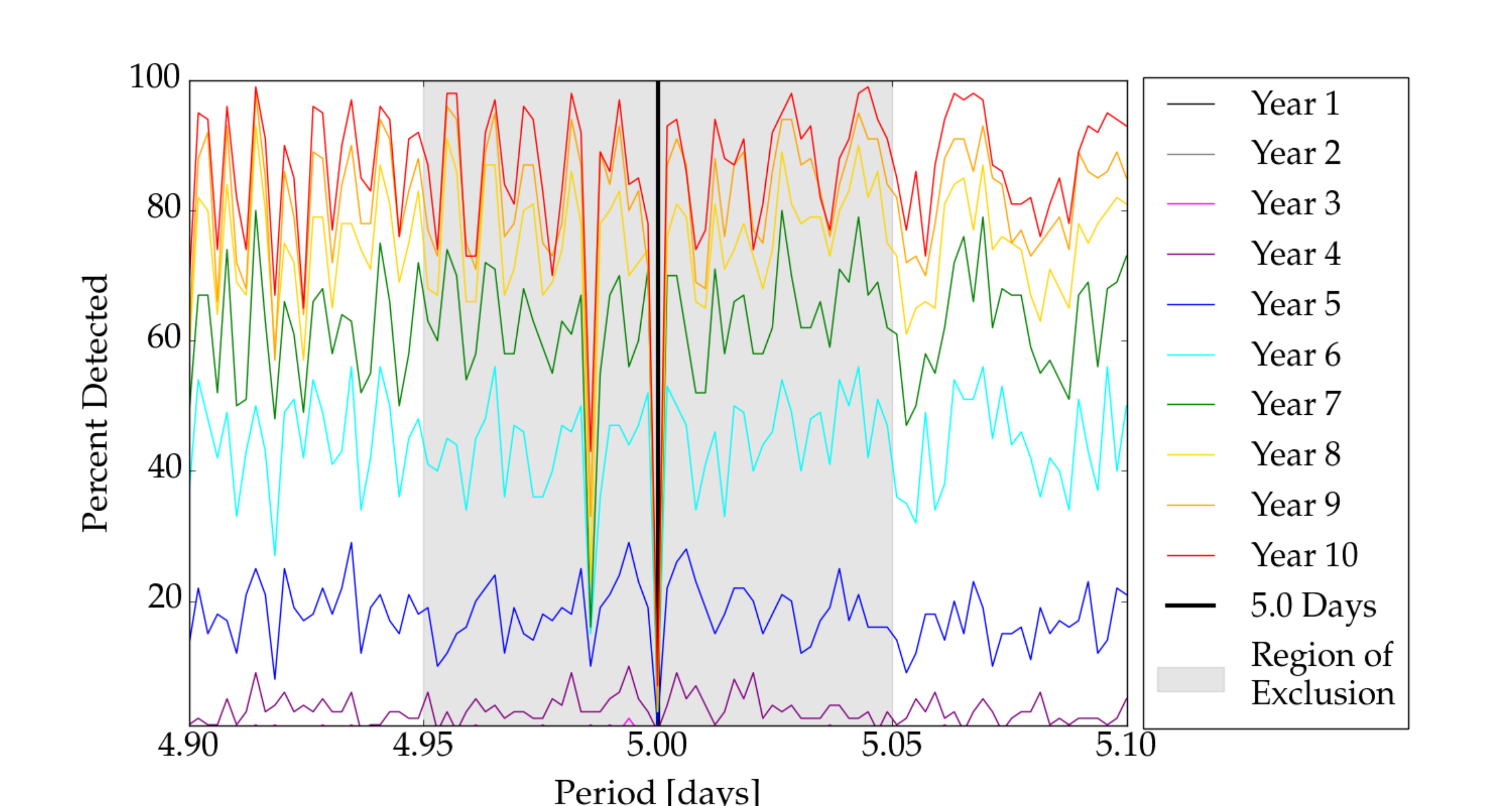}
    \caption{This figure shows a detection drop at period that is an integer multiple of 1 day ($\sim$5 days) and 1 sidereal day ($\sim4.98$ days) for a $1 M_{\odot}$, $12 R_{\oplus}$ transiting exoplanet system at 7 kpc.  Points that fall within the region of exclusion (gray area) are below the time resolution of our later simulations.  We therefore exclude parameter space within 0.05~d of integer and half integer multiples of 1~d from period searches for the remainder this study.
    }
    \label{fig:integer_day}
\end{figure}

As a ground-based telescope, exoplanet detection using LSST will necessarily be affected by diurnal observing patterns.  \citet{Jacklin:2015} showed that the effects of diurnal sampling specifically limit the detection probability at periods of integer and half-integer days.  As a check, we finely sampled the range of periods from 4.9 days to 5.1 days with a resolution of 0.01~d for a $12 R_{\oplus}$ planet orbiting a $1 M_{\odot}$ star at 7 kpc over ten years of observation.  Figure \ref{fig:integer_day} shows the detection probability, with the expected drop in detection at a period of exactly five days.  Based on the sharpness of this probability decrease we do not consider periods within 0.05d of integer and half-integer days for the remainder of this analysis.  A similar sharp decrease in detection is exhibited at an integer sidereal day (at $\sim4.98$ solar days); this feature is within 0.05d of the integer-day feature and thus is also excluded from the remainder of our analysis.

\newpage
\section{Results: Transiting Planet Detectability During the LSST 10-year Mission}\label{sec:results}

In this section, we present the resulting transiting planet detection probabilities as a function of time for each of the three fiducial cases we simulated (Section~\ref{sec:systems}), in turn.

\subsection{Hot Jupiter Detection}

\begin{figure}[!htb]
 \centering
  \includegraphics[width=\linewidth]{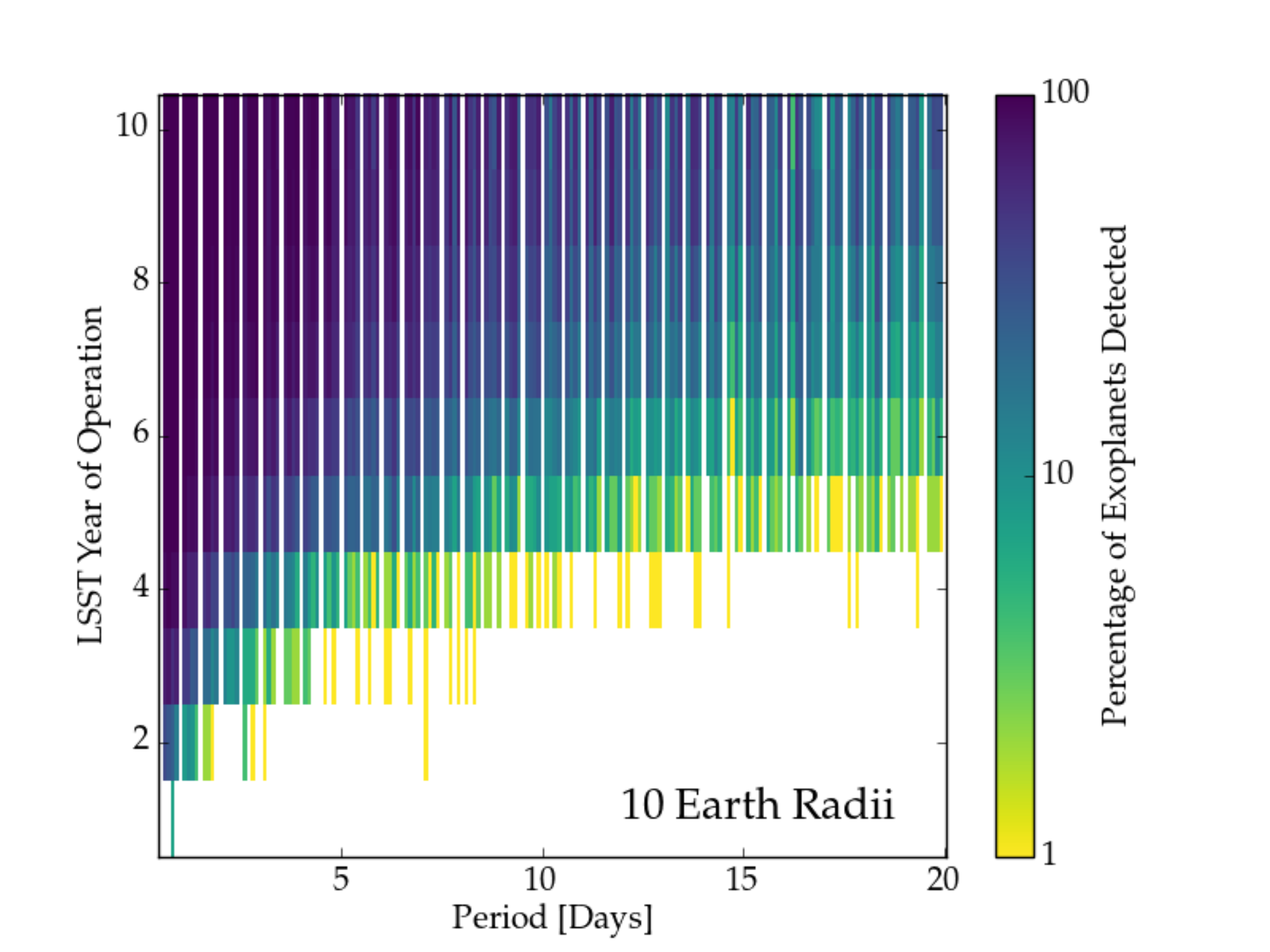}
\caption{Two-dimensional histogram across orbital period and year of observation for a G-dwarf located in a deep-drilling field.  This figure shows the results of simulating a $10 R_{\oplus}$ transiting hot Jupiter at 7 kpc.  The logarithmic color bar indicates the percent of total cases where the period of the planet is recovered to within 0.1\% accuracy with an accompanying power that crosses the power threshold for a null transit of the same system. Periods at integer and half-integer days are removed in order to mitigate the effects of diurnal sampling.}
\label{fig:gdwarf}
\end{figure}

\begin{figure}[!htb]
    \centering
    \includegraphics[width=\linewidth]{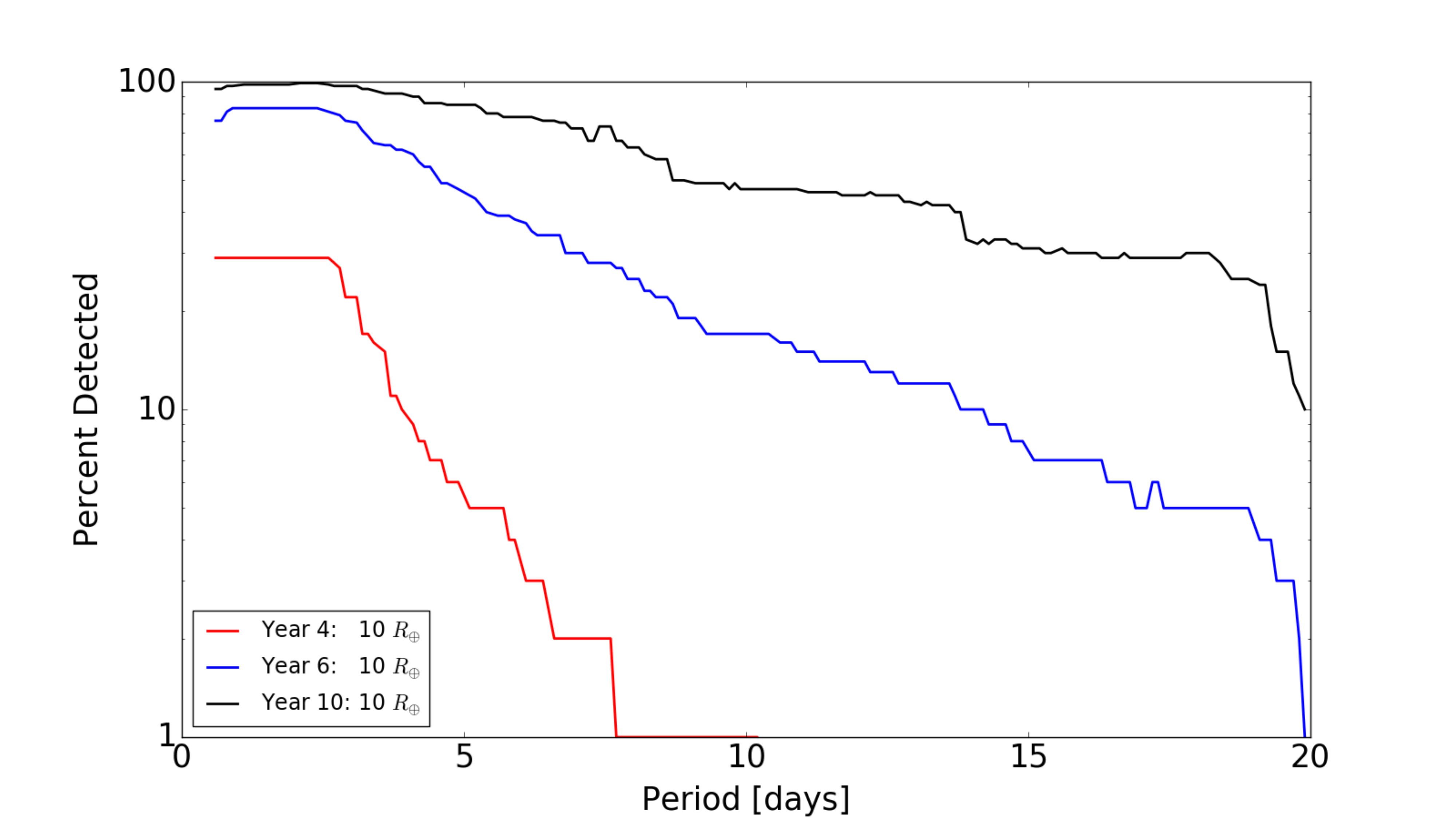}
    \caption{Detection probability as a function or orbital period for a G-dwarf with a 10 $R_{\oplus}$ transiting exoplanet at 7 kpc, based on light curves after 4, 6, and 10 years of LSST operations. The detection curves have had integer and half-integer periods removed, and are smoothed over a 3-day window.}
    \label{fig:smooth_10g}
\end{figure}

The most successful exoplanet detection in the region of parameter space tested occurred with large planets orbiting a $1 M_{\odot}$ host star at 7~kpc (see Figure~\ref{fig:systems}).  Here we tested a $10 R_{\oplus}$ transiting exoplanet (Figure~\ref{fig:gdwarf}), representing the original system analyzed for period recoverability in \citet{Jacklin:2015}.  As shown by the logarithmic shading in Figure~\ref{fig:gdwarf}, the large size of the transiting exoplanet yields a high probability of detection.  Appreciable detection at very short periods ($<$ 3~days) is seen after about 4 years of observation, with excellent detection probabilities all the way out to 20-day periods after ten full years of observation.  This overall pattern is summarized in Figure~\ref{fig:smooth_10g}.

\subsection{Hot Neptune Detection}

\begin{figure}[!htb]
  \centering
  \includegraphics[width=\linewidth]{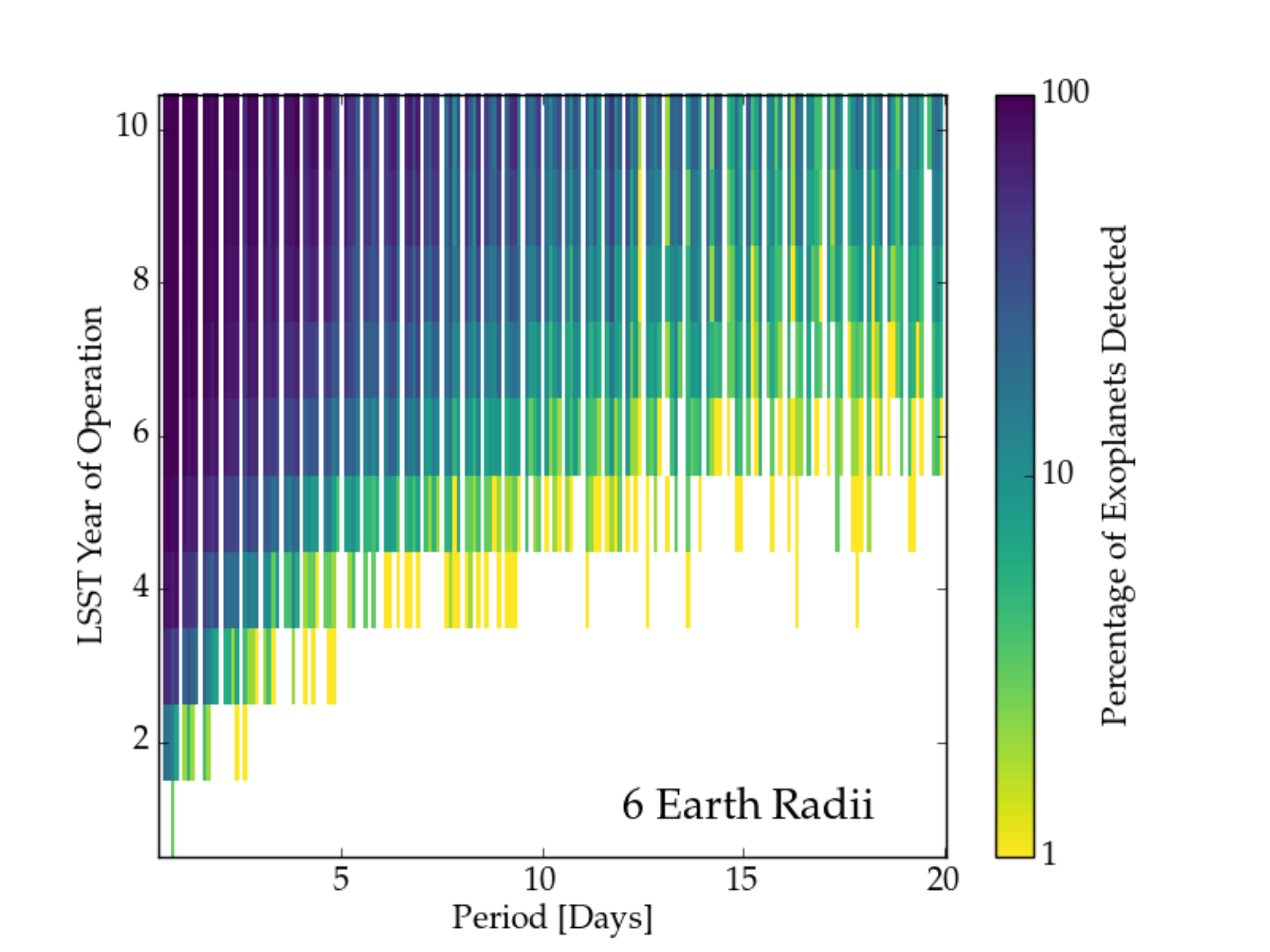}
  \includegraphics[width=\linewidth]{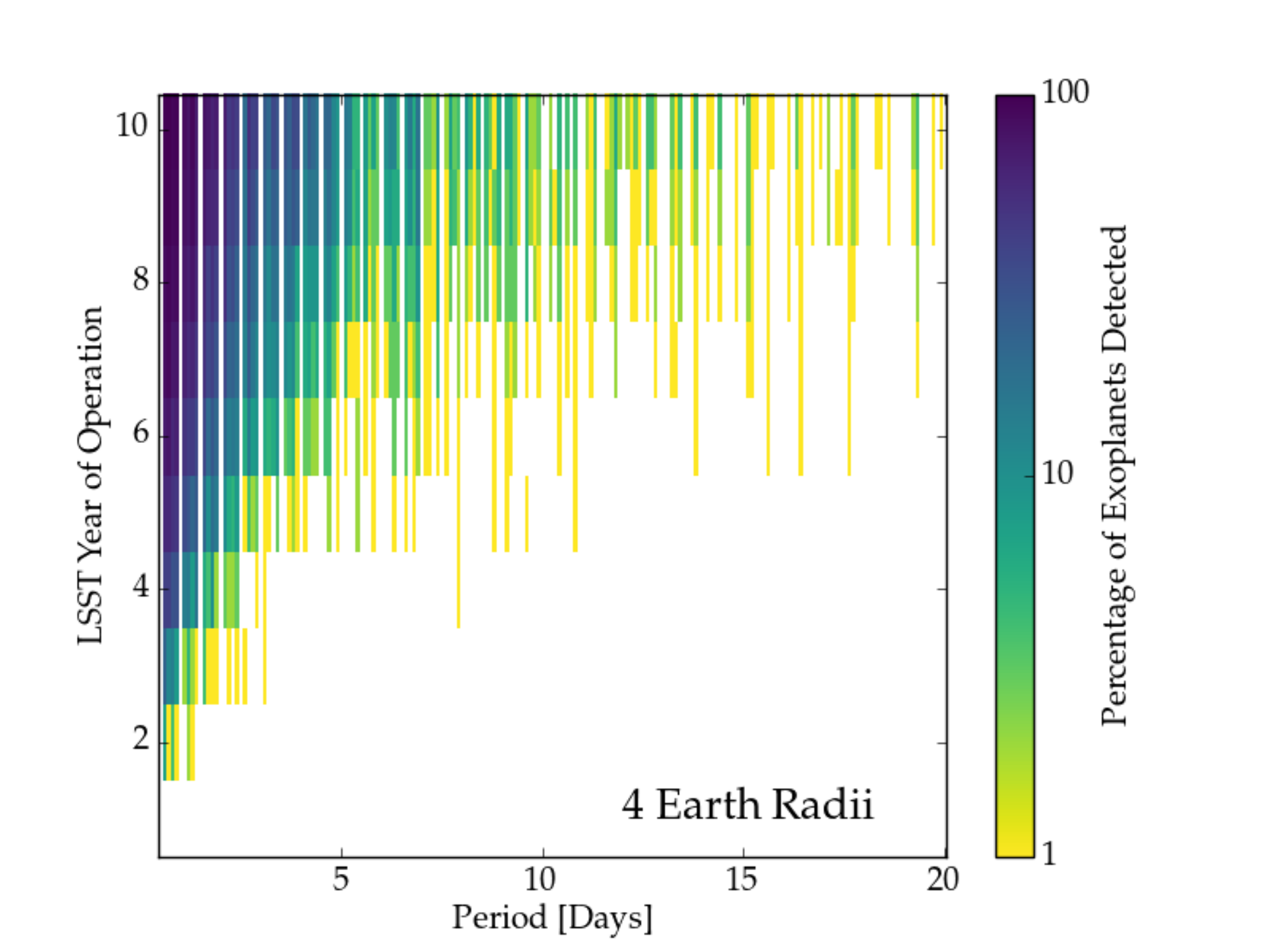}
\caption{Two-dimensional histograms across orbital period and year of observation for a K-dwarf located in a deep-drilling field.  The top figure simulates a $6 R_{\oplus}$ planet, and the bottom figure represents a $4 R_{\oplus}$ planet, both utilizing a similar legend to Figure~\ref{fig:gdwarf}. Periods at integer and half-integer days are removed in order to mitigate the negative effects of diurnal sampling.}
\label{fig:kdwarf}
\end{figure}

\begin{figure}[!htb]
    \centering
    \includegraphics[width=\linewidth]{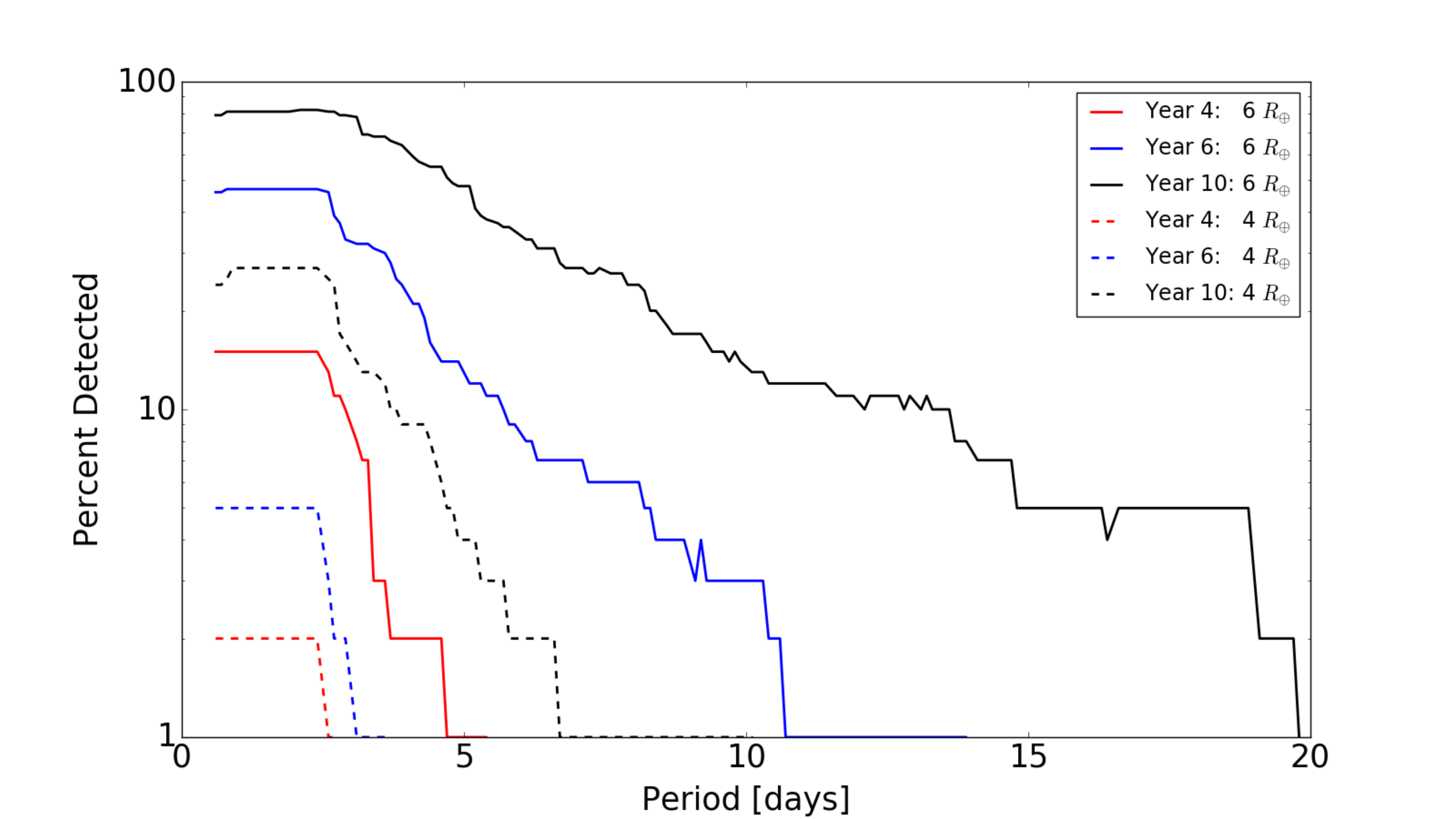}
    \caption{Detection probability as a function or orbital period for a K-dwarf at 2 kpc with a transiting exoplanet of 6 $R_{\oplus}$ (solid lines) and 4 $R_{\oplus}$ (dashed lines). The data have been processed in the same manner as Figure~\ref{fig:smooth_10g}.}
    \label{fig:smooth_6k}
\end{figure}

Results for a hot transiting Neptune-sized planet are also promising. We analyzed a $6 R_{\oplus}$ planet and a $4 R_{\oplus}$ planet transiting a $0.6 M_{\odot}$ K-dwarf at 2~kpc (Figure~\ref{fig:kdwarf}).  Detection of the $6 R_{\oplus}$ exoplanet is high out to roughly 8-day periods after seven years of observation, with detection probabilities increasing with further years of observation.  Detection of the $4 R_{\oplus}$ exoplanet is more difficult, with high detection occurring only at $<$ 3-day periods after seven years of observation.  This overall pattern is summarized in Figure~\ref{fig:smooth_6k}.

\subsection{Hot Super Earth Detection}

\begin{figure}
  \centering
  \includegraphics[width=\linewidth]{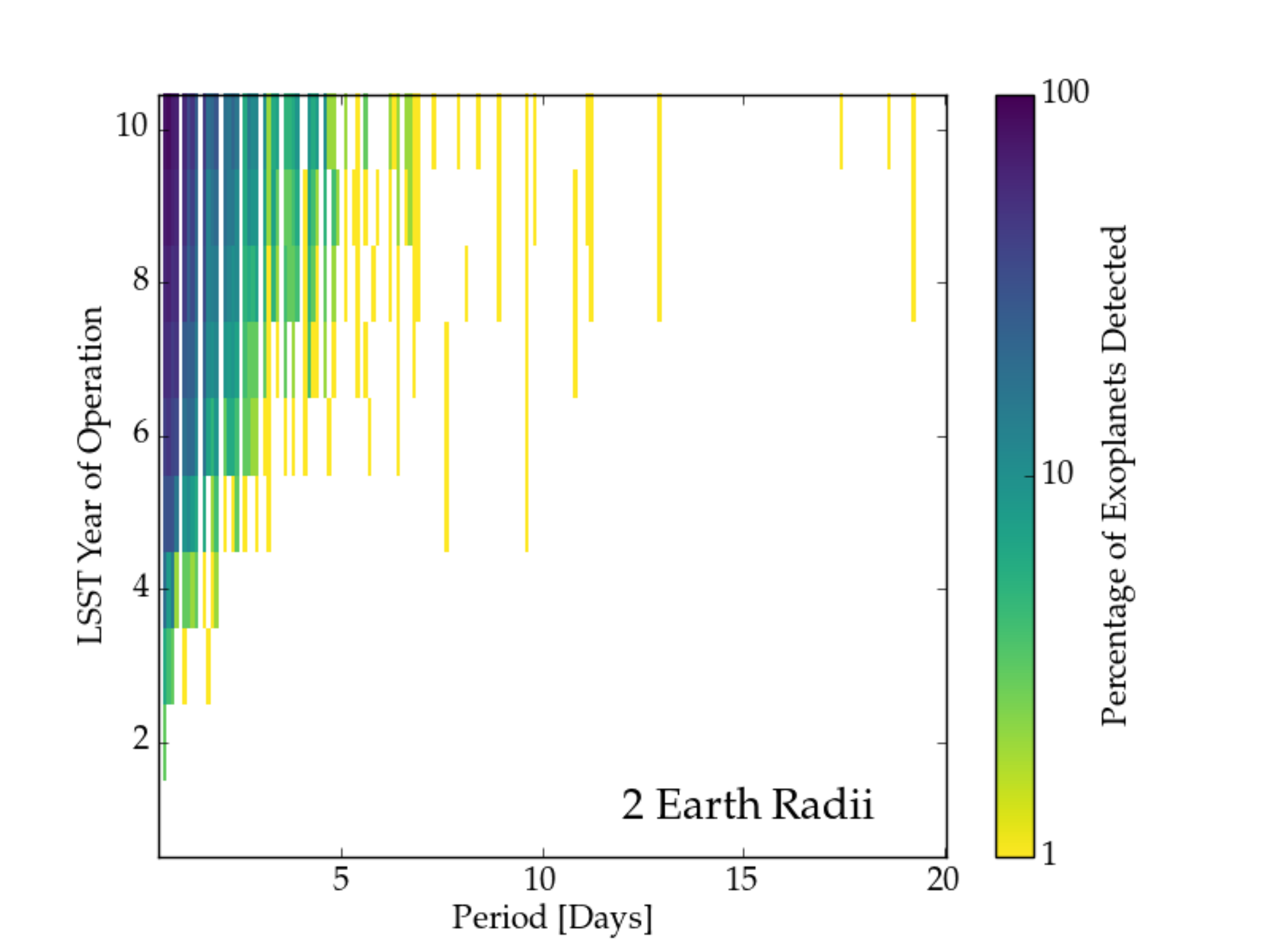}
  \caption{M-dwarf deep-drilling field two-dimensional histogram in period and year of observation space.  This figure simulates a $2 R_{\oplus}$ transiting hot Super Earth at 200 pc.  This figure was created using the same data processing used for Figure~\ref{fig:gdwarf} and Figure~\ref{fig:kdwarf}.}
  \label{fig:mdwarf}
\end{figure}

\begin{figure}
    \centering
    \includegraphics[width=\linewidth]{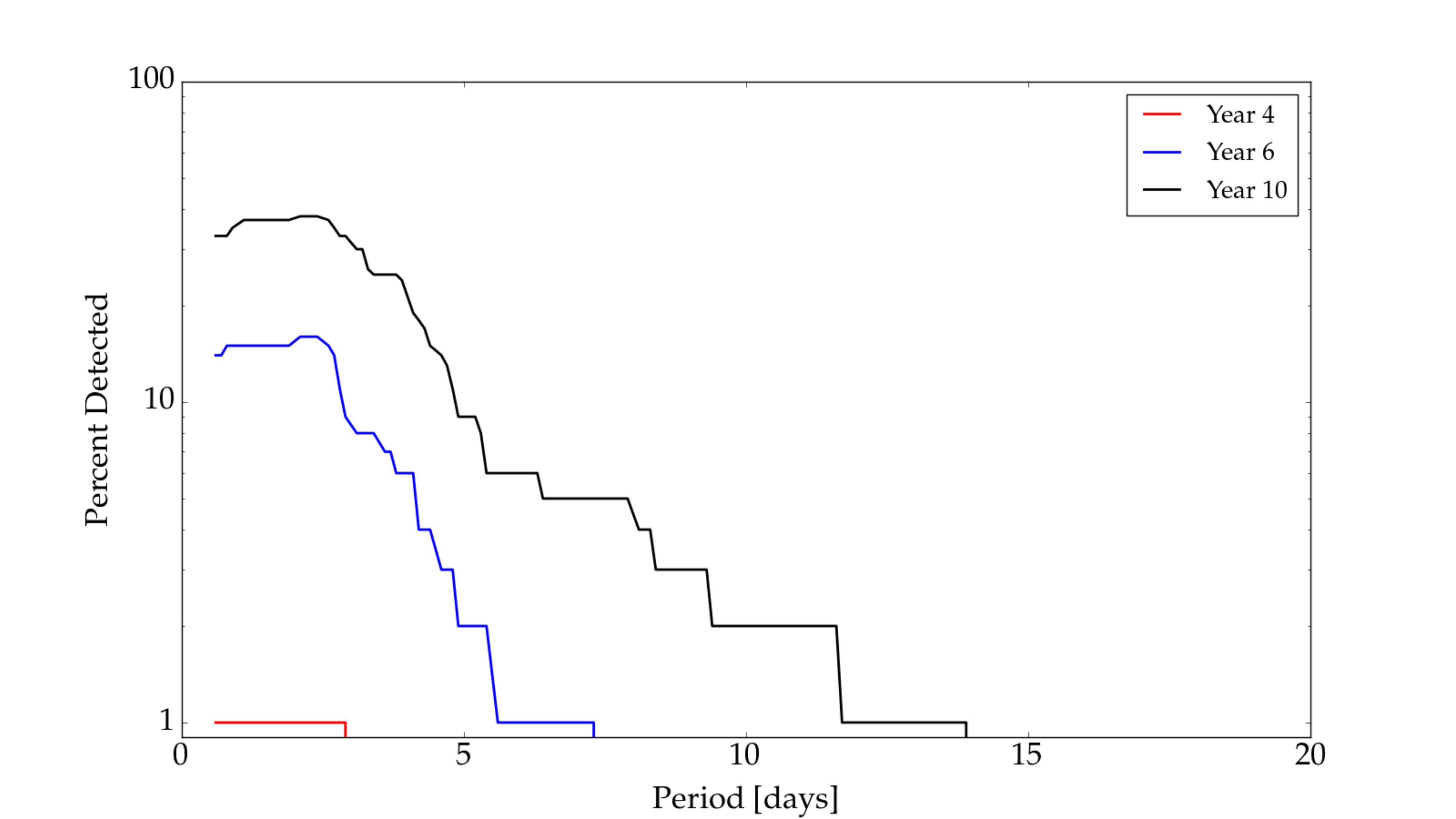}
    \caption{M-dwarf with at 2 $R_{\oplus}$ transiting exoplanet at 200 pc.The data have been processed in the same manner as Figure~\ref{fig:smooth_10g} and Figure~\ref{fig:smooth_6k}}
    \label{fig:smooth_2m4}
\end{figure}

The final system analyzed is a $2 R_{\oplus}$ transiting exoplanet orbiting a $0.25 M_{\oplus}$ M-dwarf at a distance of 200~pc.  This system is the most Earth-like of the examples considered here and could potentially represent a rocky planet in an optimistic habitable zone as mentioned by \citet{Selsis:2007} and defined by \citet{Kasting:1993} at the fringes of our exoplanet detection space as defined by number of observations in Figure~\ref{fig:systems}.  The $0.25 M_{\oplus}$ system at 200~pc represents an extreme in the LSST parameter space for transiting planet discovery as it is the closest to Earth that an $0.25 M_{\oplus}$ star can be located without nominally saturating the LSST detectors.  As shown in Figure~\ref{fig:mdwarf}, exoplanet detection in this case is difficult but not impossible.  For the tested system, there is a very high probability that we will be able to detect $2 R_{\oplus}$ exoplanets at periods shorter than $\sim3$ days after $\sim6$ years.  As the period duration increases, the detection probability drops dramatically as summarized in Figure~\ref{fig:smooth_2m4}.

\section{Discussion and Summary}\label{sec:disc}

These results are a continuation of the work of \citet{Lund:2015} and \citet{Jacklin:2015}.  We have now expanded our parameter space beyond the fiducial hot Jupiter orbiting a solar-type star to include more varied exoplanetary systems, and to test the boundaries of LSST's exoplanetary detection capabilities.  
 
Here we focus solely on LSST's deep-drilling fields, as we have found that exoplanet detection is significantly improved by using deep drilling cadence as opposed to regular cadence.  Additionally, since rapid results are of increasing importance as first light approaches, we examine not only what LSST will discover after its full tenure, but have also specifically explored exoplanet detection as LSST steadily accumulates more data.  Indeed, this work has demonstrated that at least a few percent of the shortest-period ($\lesssim2$~d) hot Jupiters transiting G-dwarf stars and hot Neptunes transiting K-dwarf stars can be detected within the first 1--2~yr of LSST data collection (Figures \ref{fig:gdwarf} and \ref{fig:kdwarf}).  More generally, detection probabilities of $\gtrsim$10\% for periods $\lesssim$5~d are possible mid-way through the nominal 10-year LSST survey.

In this paper we have simulated five fiducial test cases that broadly explore LSST's exoplanet detection parameter space.  As a next step we plan to more fully sample LSST's observational parameter space for transiting exoplanet detectability based on stellar mass, exoplanet radius, system distance, and exoplanet period for both LSST deep-drilling and regular cadence.  It will also be useful to consider the effects of eccentric orbits and grazing transits.

As expected, very short-period hot Jupiters transiting mid-type main-sequence stars (i.e., $1 M_{\odot}$) are the easiest exoplanets that LSST will be able to detect, however they will be different from the systems explored by most other current surveys due to their faintness and distance from Earth.  More generally, our work has shown that LSST will be capable of detecting a wide variety of exoplanets over a range of parameter space that is previously underexplored, including hot Jupiters at great distances and super-Earths around red dwarfs.  Specifically, LSST will have the ability to probe planets orbiting distant and/or intrinsically dim stars
($r$ $\sim24.5$) over the entire southern sky.  Finally, because the specific cadences and deep-drilling fields for LSST have not yet been fully defined, the work presented here may further help to develop optimal choices for the telescope. 

Our simulations can be enhanced in a number of ways to more fully and realistically assess the range of conditions under which these types of transiting planets may be discovered by LSST. Our treatment of noise in the LSST light curves currently neglects both the effects of contamination (e.g., crowding) and of stellar variability (e.g., activity). Future simulations could attempt to estimate likely contamination ratios for different positions on sky, as has been done for the TESS Input Catalog \citep{Stassun:2014}, and could also include a random sampling of typical activity levels as a function of stellar spectral type. This could be especially important for very late-type stars. We have also not attempted here to estimate the absolute number of transiting planets that may be discovered by LSST. To do this will require a comprehensive assessment of various types of false positives, which the additional capabilities described above would enable.  Lastly, we have also not yet grappled with the probability that most, if not all, LSST transit detections may not be possible to confirm using traditional dynamical techniques (e.g., radial-velocity followup) due to the host star faintness, so statistical methods will be required to translate such detections into more useful results. 

Most importantly, as we have demonstrated with a number of exemplar cases here, LSST should be capable of discovering a variety of exotic exoplanetary systems.  A more complete exoplanetary census will contribute to deeper understanding of the frequency, structure, and formation of these systems in our galaxy and beyond.

\acknowledgments
S.J.\ and K.G.S.\ gratefully acknowledge partial support from NSF PAARE grant AST-1358862.  S.J. also acknowledges considerable academic insight from Natalie Hinkel.

\bibliographystyle{apj}
\bibliography{paper2_bib}

\end{document}